\documentclass[aps,showpacs,pra,preprint]{revtex4}
\usepackage{mathrsfs} 
\usepackage{amssymb}

\usepackage{graphicx}
\usepackage{dcolumn}
\usepackage{bm}

\begin{document}

\title{Non-Markovian Entanglement Sudden Death and Rebirth of a Two-Qubit System in the Presence of System-Bath Coherence}
\author{Hao-Tian Wang, Chuan-Feng Li$\footnote{
email: cfli@ustc.edu.cn}$, Yang Zou, Rong-Chun Ge and Guang-Can Guo}
\affiliation{Key Laboratory of Quantum Information, University of
Science and Technology of China, CAS, Hefei, 230026, People's
Republic of China}
\date{\today }


\begin{abstract}
We present a detailed study of the entanglement dynamics of a
two-qubit system coupled to independent non-Markovian environments,
employing hierarchy equations. This recently developed theoretical
treatment can conveniently solve non-Markovian problems and take
into consideration the correlation between the system and bath in an
initial state. We concentrate on calculating the death and rebirth
time points of the entanglement to obtain a general view of the
concurrence curve and explore the behavior of entanglement dynamics
with respect to the coupling strength, the characteristic frequency
of the noise bath and the environment temperature.
\end{abstract}

\maketitle

\section{introduction}
Quantum information, including quantum entanglement and quantum
dissonance \cite{introduction1.5,introduction1,introduction3}, has
been a popular area of research in recent years. In this field,
entanglement undoubtedly plays a central role as an important
resource in quantum computation \cite{introduction1.6},
teleportation \cite{othermethod3.1}, dense coding
\cite{introduction2.1} and cryptography
\cite{othermethod3.2,othermethod3.3}. There have been many
meaningful works on the dynamics of entanglement in multiple qubits,
especially two qubits, interacting with different kinds of
environments and we now know some important features of the
entanglement such as the entanglement sudden death
\cite{othermethod4.1,othermethod2,introduction1.2,introduction1.3}
and birth \cite{othermethod3.4}.

However, environments with different properties will have different
effects on the entanglement dynamics. Recently quantum systems in a
non-Markovian environment have been a subject of great interest
\cite{first,othermethod4,othermethod1,introduction1.1,introduction1,introduction2,introduction1.4}.
Theoretical treatments have been developed to deal with this
situation, such as Ref.\,\cite{othermethod4}, \cite{othermethod1}
and \cite{othermethod3}. Of particular interest to us is the
hierarchy equations approach employed by Dijkstra and Tanimura
\cite{first}. This treatment is deduced using the influence
functional method of Feynman and Vernon without the limitation of
perturbative, Markovian or rotating wave approximations. It can
easily take into account the effect of the system-bath coupling on
the dynamics of the entanglement for any initial conditions. In
their work, they discussed the entanglement evolution of two qubits
interacting with a quantum-mechanical bath and then compared this
hierarchy method with the Redfield equation. It is found that the
result of the full calculation markedly differs from the Redfield
predictions. In the present work we continue this meaningful work.

We first introduce the method of hierarchy equations of motion
developed by Y. Tanimura and coworkers in
Ref.\,\cite{equation1,equation2,equation3}. Then we employ this
method to carefully calculate the entanglement dynamics of a
two-qubit system interacting with a non-Markovian environment,
determining the influence of the strength of the system-bath
interaction, the characteristic frequency of the bath and the
environment temperature on the time evolution of the entanglement,
especially on the sudden death and sudden birth time points.

The entanglement of the two qubits should be measured using
Wootters' concurrence \cite{othermethod2.1}:
\begin{eqnarray}
\mathcal{C}(\rho)=\max(0,\sqrt{\lambda_1}-\sqrt{\lambda_2}-\sqrt{\lambda_3}-\sqrt{\lambda_4}),
\end{eqnarray}
and
\begin{eqnarray}
\rho=\rho_{AB}(\sigma_y^A\otimes\sigma_y^B)\rho_{AB}^\ast(\sigma_y^A\otimes\sigma_y^B),
\end{eqnarray}
where $\rho_{AB}$ is the density matrix of system AB,
$\rho_{AB}^\ast$ denotes the complex conjugation of $\rho_{AB}$ and
$\sigma_y^{A(B)}$ is the Pauli matrix. $\lambda_i$ are the
eigenvalues of $\rho$ ($\lambda_1$ should be the largest
eigenvalue).

\section{The Model and Theory}

For simplicity, we set $\hbar=1$ throughout this report. We assume
that the two qubits are coupled independently to two identical baths
with the same strength. The baths have a characteristic frequency
$\gamma$, and large and small $\gamma$ indicate fast and slow baths
respectively. The energy gap of the qubits is $\varepsilon$ and the
qubits are coupled by an interaction $\zeta$. Therefore, we can
write the standard system Hamilton as \cite{first,chinese}
\begin{eqnarray}
H_S=\varepsilon(a_1^{\dag}a_1+a_2^{\dag}a_2)+\zeta(a_1^\dag+a_1)(a_2^\dag+a_2).
\end{eqnarray}
The subscripts 1 and 2 represent the two qubits and $a^\dag$ and $a$
are the creation and annihilation operators. We choose a model that
adequately represents the environment as a set of oscillators with a
coupling linear in the oscillator coordinates, having the form
\cite{leggett}
\begin{eqnarray}
H_B=\sum_j(\frac{p_j^2}{2m_j}+\frac{1}{2}m_j\omega_j^2x_j^2)
\end{eqnarray}
and
\begin{eqnarray}
 H_{SB}=-\sum_{m,j}C_{mj}(a_m^\dag+a_m)x_j,
\end{eqnarray}
where $H_B$ is the Hamilton for the bath and $H_{SB}$ is the
interaction item of the system and bath. $x_j$, $p_j$, $m_j$, and
$\omega_j$ are the coordinate, momentum, mass, and frequency of the
$j$th harmonic oscillator, respectively. $C_{mj}$ is the strength of
coupling of the $m$th qubit to the $j$th oscillator. To obtain
complete information about the effect of the environment, we
introduce the spectral density $J_m(\omega)$, which is defined as
\begin{eqnarray}
J_m(\omega) =\frac{\pi}{2}\sum_{j=1}(C_{mj}^2/m_\alpha\omega_\alpha)\delta(\omega-\omega_\alpha),
\end{eqnarray}
and in this report we assume that
\begin{eqnarray}
J_m(\omega)=\omega\eta\gamma/(\omega^2+\gamma^2)
\end{eqnarray}
which comes from the Lorentzian cutoff \cite{equation2}, and is the
same for both qubits. The the hierarchy equations approach then
gives the equation of motion for the system density matrix, which
has the form \cite{chinese,first,equation3}:
\begin{eqnarray}
\frac{d\rho_{\mathbf{n}}(t)}{dt}&=&-(i\mathcal{L}+\sum_{m=1}^2\sum_{k=0}^Kn_{mk}\gamma_k)\rho_{\mathbf{n}}(t)\nonumber-\sum_{m=1}^2((\frac{1}{\beta\gamma_0}-i\frac{1}{2})\eta-\sum_{k=0}^K\frac{c_k}{\gamma_k})[V_m,[V_m,\rho_{\mathbf{n}}(t)]]\nonumber\\
&&-i\sum_{m=1}^2\sum_{k=0}^K[V_m,\rho_{n_{mk}+1}(t)]-i\sum_{m=1}^2\sum_{k=0}^Kn_{mk}(c_kV_m\rho_{n_{mk}-1}(t)-c_k^\ast\rho_{n_{mk}-1}(t)V_m),
\end{eqnarray}
where $\mathcal{L}\rho=[H_S,\rho]$, $\gamma_0=\gamma$ (the bath
frequency), $\gamma_k=2\pi k/\beta$ (for $k\geq1$) are Matsubara
frequencies, $c_0=\frac{\eta\gamma}{2}(-i+\cot\beta\gamma/2)$,
$c_k=2\eta\gamma_0\gamma_k/\beta(\gamma_k^2-\gamma_0^2)$, $k\geq1$,
and $V_m=a_m^\dag+a_m$.

In this equation, only $\rho_\mathbf{0}$ represents the physical
system density operator, and other $\rho_\mathbf{n}$ are called
auxiliary density operators (ADOs). The subscript $\mathbf{n}$ is a
multi-index, which has $2\times(K+1)$ dimensions and can be extended
as $n_{mk}$, i.e. $n_{10}$, $n_{11}$,...,$n_{1K}$; $n_{20}$,
$n_{21}$,...,$n_{2K}$. The notation $n_{mk}\pm1$ refers to an
increase and decrease of this index. When the auxiliary density
operators are all zero, the system and bath are not coupled; when
the density operators are nonzero, we should properly take the
coupling into consideration, which will influence the entanglement
evolution dramatically. All of these values contain important
information about the coupling.

The method can be employed to calculate the dynamics of a system in
a bath from any initial state (correlated or not) in a computer
program with the truncating method developed by A. Ishizaki and Y.
Tanimura \cite{equation3}. In the present work, we investigate the
behavior of the entanglement death and rebirth time points with
respect to several influential factors mentioned above. However, the
range of all parameters should be carefully determined so as to obey
the termination approximation and ensure that the results are
sufficiently precise.

\section{The Result and Discussion}

We choose $\varepsilon=1.5\zeta$, $\eta$ from $0.4\zeta$ to $0.8\zeta$, $\gamma$ from $0.4\zeta$ to $1.1\zeta$, and $\beta\zeta$ from $2$ to $3$. Our initial state is
chosen as
\begin{eqnarray}
\rho_\mathbf{0}(0)=\left(\begin{array}{cccc}
0 & 0 & 0 & 0\\
0 & 0.5 & -0.5 & 0\\
0 & -0.5 & 0.5 & 0\\
0 & 0 & 0 & 0
\end{array}
\right)
\end{eqnarray}
in the standard product $|e_1e_2\rangle$, $|e_1g_2\rangle$,
$|g_1e_2\rangle$, $|g_1g_2\rangle$, and $e_m$ and $g_m$ are the
excited state and ground state of the $m$th qubit.

\subsection{System-Bath Interaction}

\begin{figure}[b]
\centering
\includegraphics[width= 3in]{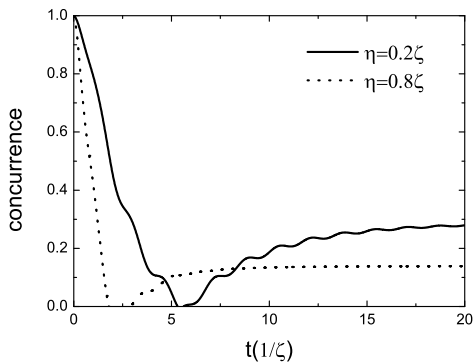}
\caption{Concurrence as a function of time with $\beta\zeta=2.5$,
$\gamma=0.5\zeta$, $\eta=0.2\zeta$ for the black line and
$\eta=0.8\zeta$ for the red line.}\label{interaction1}
\end{figure}
\begin{figure}[t]
\centering
\includegraphics[width= 3in]{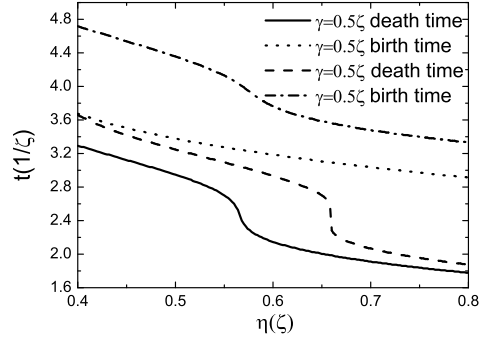}
\caption{Entanglement death time and rebirth time as functions of
$\eta$, with $\gamma=0.5\zeta$ for the the solid and dotted lines
and $\gamma=0.4\zeta$ for the dashed and dash-dotted lines,
$\beta\zeta=2.5$.}\label{interaction2}
\end{figure}

We know that $\eta$ is related to the system-bath coupling strength.
Though the function of $\eta$ in the entanglement dynamics we can
infer that when the coupling is weak, a slow decay of the
concurrence is expected because the flow of information from the
system into the bath will be slow, whereas the decay will definitely
be rapid when the system and bath are tightly correlated.
Figure\,\ref{interaction1} proves our inference; we see that the
decay of occurrence is slower when $\eta$ is small than when $\eta$
is large. The figure also indicates that small $\eta$ may result in
strong vibration of the concurrence. The environment can return a
small amount of information to the system during the evolution. With
weak coupling (small $\eta$), this little returned information will
have a strong effect on the curve of the entanglement when the
concurrence is small, and thus, there is oscillation. As $\eta$
increases, the curve becomes smoother, and there is only one pair of
death and birth time points. From Fig.\,\ref{interaction2}, we see
clearly the change in death and birth time points with the increase
in $\eta$. The negative slopes of these curves are consistent with
the effect of $\eta$ discussed above. It is interesting to find that
at some intervals the slopes of these curves suddenly change in
Fig.\,\ref{interaction2}. This phenomenon is due to a change in the
slope near the death and birth points on the concurrence curve.

\subsection{Characteristic Frequency of the Bath}

\begin{figure}[tbph]
\centering
\includegraphics[width= 3in]{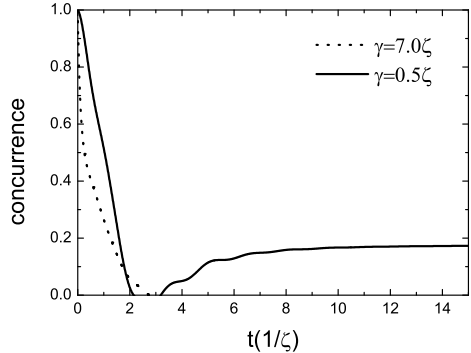}
\caption{Concurrence as a function of time with $\beta\zeta=2.5$,
$\eta=0.6\zeta$, $\gamma=7\zeta$ for the dotted line and
$\gamma=0.5\zeta$ for the solid line.}\label{bathspeed1}
\end{figure}

\begin{figure}[tbph]
\centering
\includegraphics[width= 3in]{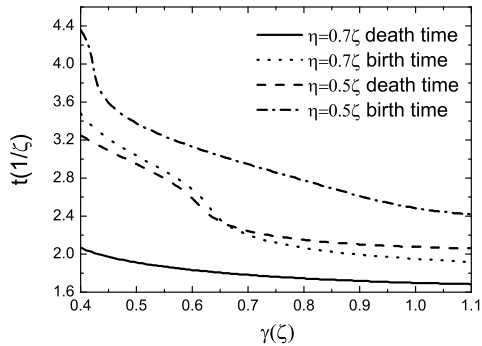}
\caption{Entanglement death time and rebirth time as functions of
$\gamma$, with $\eta=0.5\zeta$ for the dash and dash-dotted lines
and $\eta=0.7\zeta$ for the solid and dotted lines,
$\beta\zeta=2.5$.}\label{bathspeed2}
\end{figure}

Figures\,\ref{bathspeed1} and \,\ref{bathspeed2} show the influence
of the characteristic frequency of the bath on the dynamics of the
entanglement. Large $\gamma$ means a fast bath. Because the
frequency of the two level system in this report is $1.5\zeta$, and
if we take into account the range of $\gamma$ discussed here we know
that the bath is slow. Most theoretical treatments with which we
describe this entanglement dynamics are valid only if there is a
fast bath, which has a much larger characteristic frequency than the
system. Only in the fast bath scenario can we choose the initial
state in which the system and bath are not correlated \cite{first}.
However, as mentioned before, the hierarchy method can easily solve
a slow bath situation, taking into consideration the bath effect on
the entanglement in the initial state. A slow bath can receive and
return the information from the system slowly and a fast bath may
absorb all the information in a very short time and there may not
even be revival, as we show in Fig.\,\ref{bathspeed1}. This can
explain the delay in the death and birth when $\gamma$ is small.
More details are shown in Fig.\,\ref{bathspeed2}. The death and
birth time points approach zero with an increase in $\gamma$. With
smaller $\eta$, we see that all time points increase, for the reason
discussed above.

\subsection{Temperature}

\begin{figure}[tbph]
\centering
\includegraphics[width= 3in]{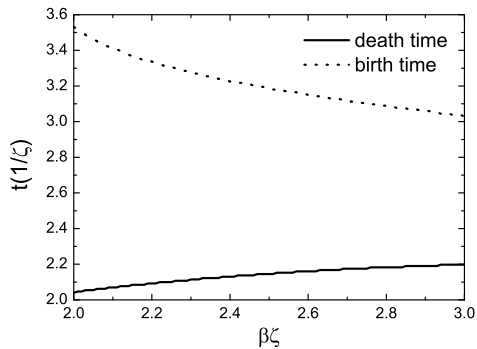}
\caption{Sudden death time point (solid line) and sudden birth time
point (dotted line) of entanglement with different values of
$\beta\zeta$ from 2 to 3. Other parameters are $\eta=0.6\zeta,
\gamma=0.5\zeta$.}\label{temperature}
\end{figure}

Temperature influences the system dramatically. In
Ref.\,\cite{first} it is seen that the low-temperature scenario is
unsuitable for Born and ultrafast bath approximations, which are
used in many theoretical treatments; however, this scenario is in
the solvable range of the hierarchy equations approach. In
Fig.\,\ref{temperature}, the death time point increases and birth
time point decreases with a decrease in temperature. That is to say,
the entanglement revives more quickly at lower temperature. The
figure infers that the concurrence curve should be sufficiently
smooth, and with the increase in $\beta\zeta$, the whole curve is
raised up to obtain a smaller death time interval. If we simulate
the two curves in the figure and calculate the relationship between
$\beta\zeta$ and the time point, then we may simply change the
environment temperature to prepare our desired states and dynamics
of the entanglement in experiments.

In fact, with higher hierarchy of the ADOs, we can investigate a
larger $\beta\zeta$ scenario. However, the computation time is too
great as it increases exponentially with higher orders.

\section{conclusion}

In this report we studied the entanglement dynamics of two-qubit in
a non-Markovian environment using the recently developed hierarchy
equations approach. We explored in detail the role of parameters in
the entanglement evolution, including the system-bath interaction
$\eta$, the temperature $\beta$ and the bath frequency $\gamma$. All
these important features are useful when preparing a desired system
state in an experiment. We also showed this hierarchy equations
approach to be effective in calculating the dynamics of a system in
many possible scenarios. Applying a higher order of the equation for
a broader range of parameters and the physical meaning of non-zero
ADOs will be future work.

\section{acknowledgement}
We appreciate helpful comments on the manuscript made by Hong-Yuan
Yuan and thank Min-Jie Zhu and Feng-Cheng Wu for modification of the
calculation program. This work was supported by the National
Fundamental Research Program and National Natural Science Foundation
of China (Grant Nos. 60921091, 10874162 and 10734060).

\end{document}